\documentclass[12pt]{article}
\usepackage[dvips]{graphicx}
\usepackage{amssymb}
\usepackage{amsmath}
\usepackage{epsfig}
\usepackage{cite}
\usepackage{hyperref}
\usepackage{bbold,bm}
\usepackage{multirow}

\def\beq{\begin{equation}}
\def\eeq{\end{equation}}
\def\be{\begin{equation}}
\def\ee{\end{equation}}
\def\bea{\begin{eqnarray}}
\def\eea{\end{eqnarray}}

\DeclareRobustCommand{\SkipTocEntry}[4]{}
\RequirePackage{color}
\newcommand{\dd}{\mathrm{d}}

\def\w{\omega}
\def\t{\tau}
\def\r{\rho}
\def\Om{\Omega}
\def\a{\alpha}
\def\ep{\epsilon}

\def\zet          {{\mathbb Z}}

\begin{document}

\phantom{AAA}
\vspace{-20mm}

\begin{flushright}

%IPHT-T19/XXX\\
CPHT-RR058.102019

\end{flushright}

\vspace{4mm}

\begin{center}

{\huge {\bf Throat destabilization}}

\medskip

{\LARGE {\bf (for profit and for fun)}}

\bigskip

\vspace{5mm}

{\large
\textsc{Iosif Bena$^1$,~Alex Buchel$^{2,3}$,~ Severin L\"ust$^{1,4}$}}

\vspace{8mm}

$^1$Institut de Physique Th\'eorique, Universit\'e Paris Saclay,\\
CEA, CNRS, F-91191 Gif sur Yvette, France \\
\medskip
$^2$Departments of Applied Mathematics, Physics and Astronomy\\ 
University of Western Ontario, London, Ontario N6A 5B7, Canada\\
\medskip
$^3$Perimeter Institute for Theoretical Physics, 
Waterloo, Ontario N2J 2W9, Canada\\
\medskip
$^4$CPHT, CNRS, Ecole Polytechnique, IP Paris, F-91128 Palaiseau, France

\vspace{4mm} 
{\footnotesize\upshape\ttfamily iosif.bena @ ipht.fr, ~ abuchel @ perimeterinstitute.ca, ~severin.luest @ polytechnique.edu} \\

\vspace{15mm}
 
\textsc{Abstract}

\end{center}

%\begin{adjustwidth}{15mm}{15mm} % to adjust the L and R margins
 
%\begin{abstract}
\vspace{-1mm}
\noindent
Two recent results indicate that the addition of supersymmetry-breaking ingredients can destabilize the Klebanov-Strassler warped deformed conifold throat. The first comes from an analytic treatment of the interaction between anti-D3 branes and the complex-structure modulus corresponding to the deformation of the conifold  \cite{Bena:2018fqc}. The second comes from the numeric construction of Klebanov-Strassler black holes  \cite{Buchel:2018bzp}, which stop existing above a certain value of the non-extremality. We show that in both calculations the destabilization energies have the same parametric dependence on $g_s$ and the conifold flux, and only differ by a small numerical factor. This remarkable match confirms 
that anti-D3 brane uplift can only work in warped throats that have an ${\mathcal O}(1000)$ contribution to the D3 tadpole.

\vspace{1.8cm}

\centerline{\it Dedicated to the memory of Steve Gubser}

%\end{abstract}
%\end{adjustwidth}

%\end{titlepage}
\thispagestyle{empty}
\newpage

%%%%%%%%%%%%%%%%%%%%%%%%%%%%%%%%%%%%%

\baselineskip=14.7pt
\parskip=3pt

\setcounter{tocdepth}{2}
%\tableofcontents
%\newpage

\baselineskip=15pt
\parskip=3pt

%\newpage

%%%%%%%%%%%%%%%%%%%%%%%%%%%%%%%%%%%%%

\noindent Adding antibranes to warped throats of flux compactifications is the preferred method to uplift the negative cosmological constant of supersymmetric flux compactifications and obtain de Sitter space in string theory \cite{Kachru:2003aw}. In previous work  \cite{Bena:2018fqc}, two of the authors together with Gra\~na and Duda\c s have shown that antibranes added to the Klebanov-Strassler warped deformed conifold (KS) solution \cite{Klebanov:2000hb} can destabilize the complex structure modulus corresponding to the size of the three-cycle of the deformed conifold, thus collapsing the Klebanov-Strassler warped deformed conifold to the Klebanov-Tseytlin (KT) singular warped conifold solution \cite{Klebanov:2000nc}. 

The destabilization mechanism found in  \cite{Bena:2018fqc} is triggered when the number of antibranes is larger than a certain bound that depends on the RR-flux $M$ on the three-cycle of the deformed conifold. Furthermore, this happens both when the KS solution is infinite, and when the KS solution is glued to a compact space. This result has led the authors of  \cite{Bena:2018fqc} to conjecture the existence of a KS black hole. The intuition behind this conjecture was based on the fact that both antibranes and a black hole horizon break supersymmetry by
making the mass of the solution larger than the charge. Hence, if a small number of antibranes does not destabilize the three-cycle of the deformed conifold when $M$ is large enough, neither one would expect a small mass above extremality brought about by a finite-temperature event horizon to do it.

In parallel to this work, one of the authors had succeeded in constructing numerically the Klebanov-Strassler black hole \cite{Buchel:2018bzp}, despite prevailing expectations based on previous work \cite{Buchel:2010wp} that this black hole should not exist and any attempt to add an event horizon to the KS solution would shrink the size of the three-cycle (or alternatively reduce the vev of the operator parameterizing chiral symmetry breaking to zero) and result in the Klebanov-Tseytlin black hole \cite{Aharony:2007vg}. Interestingly enough, the KS black hole found in \cite{Buchel:2018bzp} also exists only when the energy above extremality is below a certain value. When the energy is above this value chiral symmetry is restored and the only possible solution is the KT black hole.

The purpose of this note is to compare the (numerically found) energy above extremality at which the numerically constructed KS black hole stops existing with the (analytically found) energy above extremality at which a configuration of antibranes in a KS throat stops existing. At first glance one could argue that these two energies should have nothing to do with each other, as one energy is calculated using a fully-backreacted black-hole solution in which the nonlinear effect of the mass above extremality has been taken into account, while the other is calculated in a probe approximation which ignores the backreaction of the antibranes. Furthermore, it is well known that the KS black hole suffers from Gregory-Laflamme-like instabilities corresponding to clumping along the field theory directions \cite{Buchel:2018bzp}, while antibranes also have tachyons  on their worldvolume which make them unstable to moving away from each other on the Coulomb branch \cite{Bena:2014jaa,Bena:2016fqp}.

One can however argue that the instabilities of the KS black hole and antibranes, while important for holography and for flux compactifications, are not of direct relevance here. What we are comparing are the energies at which the two non-extremal configurations stop existing as solutions to the equations of motion, and this comparison is meaningful because the underlying phenomenon is the same: the size of the $S^3$ of the deformed conifold, or alternatively the vev of the field dual to chiral symmetry breaking is shrunk by the addition of mass above extremality. 

The comparison of the numerical results of  \cite{Buchel:2018bzp} with the analytical results of  \cite{Bena:2018fqc} is easiest to illustrate if one converts the energy above extremality of the black hole in ``antibrane'' units, corresponding to the energy above extremality brought about by a single antibrane placed in the warped deformed conifold. The KS black hole solution exists when 
\begin{equation}
g_s M^2~ \geq ~\gamma_{BH}^2 N_{\overline{D3}} %~\approx 17.3 ~ N_{\overline{D3}} 
\,,
\label{limit-BH}
\end{equation}
while the antibranes do not have a runaway when 
\begin{equation}
g_s M^2 ~\geq ~\gamma_{\overline{D3}}^2 N_{\overline{D3}} %~\approx 144 ~N_{\overline{D3}} 
\,.
\label{limit}
\end{equation}
In this paper we evaluate $\gamma_{BH}$ and $\gamma_{\overline{D3}}$ and find them to be 
\begin{equation}
\gamma_{BH}  \approx 4.16~~~~~~~~~{\rm and} ~~~~~~~~~~\gamma_{\overline{D3}} \approx 6.8 \,.
\label{gamma}
\end{equation}
This match is quite remarkable, both because the functional expression of the critical energies as functions of the parameters of the solution are the same, and because the coefficients differ by so little despite the fact that we are comparing two very different sources of non-extremality. Furthermore, one should not forget that the value of $ \gamma_{\overline{D3}} $ found in \cite{Bena:2018fqc} is based on the analysis of \cite{Douglas:2008jx}, in which the values of certain numerical coefficients were only estimated and not rigorously computed. Hence, this value also has error bars. 

This remarkable match strongly reinforces the conclusion of  \cite{Bena:2018fqc}, that the only warped KS throats in flux compactifications that are not destabilized by the addition of a single antibrane must have a RR three-form flux, $M$, larger then the value given in equation \eqref{limit}. If one is to build a de Sitter flux compactification with stable moduli, the length of this throat also needs to be quite large, which puts a lower limit on the amount of the NS-NS three-form flux on the B-cycle. This in turn implies that the contribution of the fluxes of this KS throat to the D3 tadpole of the compactification is larger than about $1000$ \cite{Bena:2018fqc}. 

Hence our analysis greatly reduces the space of Calabi-Yau manifolds where one may hope to build a de Sitter solution in String Theory, by eliminating from the landscape all the manifolds whose geometrical structure does not allow a negative contribution to the tadpole whose absolute value is smaller than about $1000$. In an upcoming paper \cite{tadpole-paper} two of the authors and collaborators will show that most of these manifolds have unfixed moduli, and reduce the question of the existence of de Sitter flux compactifications to a rigorous problem that can be prover or disproven mathematically.

\newpage
%\bigskip

\noindent $\bullet$ {The Calculation}

\bigskip

\noindent The (Einstein-frame) ten-dimensional metric and five-form flux of the Klebanov-Strassler solution \cite{Klebanov:2000hb}  take the form
\begin{equation}\begin{aligned}\label{eq:warpedbackground}
\dd s^2 &= H^{-1/2} d s^2_{4}+ H^{1/2} d s^2_{6} \,, \\
F_5 &= \left(1+\ast\right) \mathrm{vol}_4 \wedge d H^{-1}  \,,
\end{aligned}\end{equation}
where $d s^2_{6}$ is the metric of the deformed conifold.
For $M$ units of RR flux on the compact three-cycle of the deformed conifold, 
\begin{equation}
\frac{1}{4 \pi^2 \alpha'} \int_{S^3} F_3 = M \,,
\label{defm}
\end{equation}
the warp factor is given by
\begin{equation}\label{eq:warpfactor}
H({\tau})= 2^{2/3} \frac{{g_s (\alpha' M)^2}}{\left|S\right|^{4/3}} I({\tau}) \,,
\end{equation}
with $\tau$ a radial coordinate on the deformed conifold and $S$ its deformation
parameter ($|S|\equiv \epsilon^2$ in \cite{Klebanov:2000hb}), corresponding to the size of the three-sphere at the tip of the cone. In Appendix~\ref{appendixA} we outline this solution in more detail, and explain some of the subtleties related to its expressions in String frame and Einstein frame. The function $I(\tau)$ is given by an integral expression, here we will only need its numerical value at the tip of the throat,
\begin{equation}
I(0) \approx 0.718 \,.
\end{equation}

The black hole numerically constructed in \cite{Buchel:2018bzp} is stable as long as its energy density satisfies
\begin{equation}
\mathcal{E} \leq \mathcal{E}_{\chi SB} \,, 
\end{equation}
for larger values the chiral symmetry is restored \cite{Buchel:2010wp}.
After introducing
\begin{equation}\label{eq:chiSB}
\hat{\mathcal{E}} \equiv 216\pi^4(\a')^4\ \mathcal{E} \,,
\end{equation}
the value of $\hat{\mathcal{E}}_{\chi SB}$ was numerically determined in \cite{Buchel:2018bzp},
\begin{equation}\label{eq:chiSBnum}
\hat{\mathcal{E}}_{\chi SB}  = 1.270093(1) \Lambda^4 \,.
\end{equation}
Here, $\Lambda$ denotes the strong coupling scale, given by
\begin{equation}
\Lambda = \frac{3^{1/2}e^{1/3}|S|^{1/3}}{2^{7/12}} \,.
\label{deflambda}
\end{equation}

Following \cite{Bena:2018fqc} the energy density of an anti-D3-brane placed in the (Einstein-frame) Klebanov-Strassler background can be computed from its DBI and Chern-Simons action,
\begin{equation}\begin{aligned}\label{eq:D3action}
S_{D3} = S_{DBI} + S_{CS}
= - {T_3}\int \dd^4 x \sqrt{-g_4} - T_3 \int C_4 \,,
\end{aligned}\end{equation}
where its tension is given by
\begin{equation}
T_3 = \frac{1}{(2 \pi)^3 \alpha'^2} \,.
\end{equation}
Using \eqref{eq:warpedbackground} the energy density of $N_{\overline{D3}}$ anti-D3-branes, placed at the tip of the KS-throat, is therefore given by
\begin{equation}
\mathcal{E}_{\overline{D3}} = 2 N_{\overline{D3}} \, T_3 C_4 =  \frac{2 N_{\overline{D3}}}{(2 \pi)^3  \alpha'^2} H(0)^{-1} \,.
\end{equation}
Using the warp factor \eqref{eq:warpfactor} we obtain
\begin{equation}
\hat{\mathcal{E}}_{\overline{D3}} = N_{\overline{D3}} \frac{3^3 2^{1/3} \pi}{I(0)} \frac{\left|S\right|^{4/3}}{g_s  M^2} \,,
\end{equation}
and thus
\begin{equation}
\frac{\hat{\mathcal{E}}_{\overline{D3}}}{\Lambda^4} = \beta \frac{N_{\overline{D3}}}{g_s M^2} \,,
\end{equation}
where we abbreviated the numerical prefactor,
\begin{equation}\label{eq:D3num}
\beta \equiv \frac{2^{8/3} 3 \pi}{e^{4/3} I(0)}  \approx 22.0 \,.
\end{equation}
If we now naively replace in \eqref{eq:chiSB} the energy density $\mathcal{E}$ of the black hole with the  energy density of the anti-D3-branes, we find a bound of the same functional form as the one derived in \cite{Bena:2018fqc},\footnote{The factor of $g_s$ was corrected in \cite{Blumenhagen:2019qcg}.}
\begin{equation}
g_s M^2 \geq \gamma_{BH}^2 N_{\overline{D3}}  \,,
\end{equation}
where combining \eqref{eq:chiSBnum} and \eqref{eq:D3num} gives
\begin{equation}
\gamma_{BH} \approx 4.16 \,.
\end{equation}
This result is of the same order of magnitude as the
value $\gamma_{\overline{D3}} \approx 6.8 $ determined in \cite{Bena:2018fqc} by completely different methods.

Appendix~\ref{appendixA} explains in detail the KS geometry and how to do holography therein. Appendix~\ref{appendixB} explains how to do holography in the black hole solution of \cite{Buchel:2018bzp} and how to obtain equations \eqref{eq:chiSB} and \eqref{deflambda} which are crucial for relating the two calculations.

\medskip

\noindent We would like to thank Centro de Ciencias de Benasque for the hospitality
when this project started. 
The work of IB is partially supported by the ANR grant Black-dS-String ANR-16-CE31-0004-01 and the John Templeton Foundation grant 61169. Research at Perimeter
Institute is supported by the Government of Canada through Industry
Canada and by the Province of Ontario through the Ministry of
Research \& Innovation. The research of AB is further supported by
NSERC through the Discovery Grants program.
The work of S.L.\ is supported by the ERC Starting Grant 679278 Emergent-BH.

%%%%%%%%%%%%%%%%%%%%%%%%%%%%%%%%%%%%%%%%%%%%%%%%%%%%

%%%%%%%%%%%%%%%%%%%%%%%%%%%%%%%%%%%%%%%%%%%%%%%%%%%%%%%%%%%

%\newpage

\appendix
\noindent

\section{KS geometry}\label{appendixA}
We review here the relevant details of the extremal supersymmetric Klebanov-Strassler geometry \cite{Klebanov:2000hb}
following \cite{Herzog:2001xk,Buchel:2013dla}.
This would allow to precisely define the strong coupling scale $\Lambda$ \eqref{deflambda}
of the dual cascading gauge theory used in the thermodynamic analysis of the model in 
\cite{Aharony:2007vg,Buchel:2010wp,Buchel:2018bzp}. Additionally, we introduce the order parameter $\xi$
(see Appendix \ref{appendixB})
for the chiral symmetry breaking in the cascading gauge theory and its geometrical counterpart representing
the relative sizes of the $2-$ and $3-$cycles of the conifold.

The ${\cal N}=1$ supersymmetric Klebanov-Strassler solution (in Einstein frame)  takes the form:
\beq\begin{aligned}\label{ks1}
ds_{10}^2=&ds_5^2+dY_5^2\,,\qquad 
ds_5^2=H_{KS}^{-1/2} \left(-dt^2+d\bm{x}^2\right)+H_{KS}^{1/2} \w_{1,KS}^2 d\t^2\,,\\
dY_5^2=&\Om_1^2 g_5^2+\Om_2^2\
[g_3^2+g_4^2]+\Om_3^2 [g_1^2+g_2^2]\,,\\
\Omega_i=&\w_{i,KS}\ H^{1/4}_{KS}\,,\qquad h_i=h_{i,KS}\,,\qquad  e^\Phi=g\,, 
\end{aligned}
\eeq
with
\beq\begin{aligned}
&h_{1,KS}=Pg_s\ \frac{\cosh \t-1}{18\sinh \t}
\left(\frac{\t\cosh \t}{\sinh \t}-1\right)\,,\qquad 
h_{2,KS}=\frac{P}{18}\left(1-\frac {\t}{\sinh \t}\right)\,,\\
&h_{3,KS}=Pg_s\ \frac{\cosh \t+1}{18\sinh \t}
\left(\frac{\t\cosh \t}{\sinh \t}-1\right)\,,
\qquad g=g_s\,,\\
&\w_{1,KS}=\frac{\epsilon^{2/3}}{\sqrt{6}{\hat K_{KS}}}\,,\qquad 
\w_{2,KS}=\frac{\epsilon^{2/3}{\hat K_{KS}}^{1/2}}{\sqrt{2}}\cosh\frac \t2\,,\qquad  \w_{3,KS}=\frac{\epsilon^{2/3}
{\hat K_{KS}}^{1/2}}{\sqrt{2}}
\sinh\frac \t2\,,\\
&{\hat K_{KS}}=\frac{(\sinh (2\t)-2\t)^{1/3} }{2^{1/3}\sinh \t}\,,\ H'_{KS}=\frac{16((9 h_{2,KS}-P)h_{1,KS}-9 h_{3,KS} h_{2,KS})}
{9\epsilon^{8/3}{\hat K_{KS}}^2\sinh^2 \t }\,,
\label{ks2}\end{aligned}
\eeq
where the radial coordinate $\t\in [0,+\infty]$ and the parameter $P$ is related to the quantized RR flux \eqref{defm} as  
\beq
P=\frac92\ M\a'\,.
\label{ks3}
\eeq
Here $g_i$ are the usual 1-forms on the deformed conifold defined as
\bea
&&g_1=\frac{\a^1-\a^3}{\sqrt 2}\,,\ \ g_2=\frac{\a^2-\a^4}{\sqrt 2}\,,\ \ 
g_3=\frac{\a^1+\a^3}{\sqrt 2}\,,\ \ g_4=\frac{\a^2+\a^4}{\sqrt 2}\,,\  \ g_5=\a^5\,,
\eea
where 
\beq\begin{aligned}
&\a^1=-\sin\theta_1 d\phi_1\,,\qquad \a^2=d\theta_1\,,\qquad \a^3=\cos\psi\sin\theta_2 d\phi_2-\sin\psi d\theta_2\,,\\
&\a^4=\sin\psi\sin\theta_2 d\phi_2+\cos\psi d\theta_2\,,\qquad \a^5=d\psi+\cos\theta_1 d\phi_1+\cos\theta_2 d\phi_2\,,\end{aligned}
\eeq
with $0\le \psi \le 4\pi$, $0\le \theta_i\le \pi$ and $0\le \phi_i\le 2\pi$.

Topologically, the compact manifold $Y_5$ is an $S^2$ fibration over $S^3$. In the deep infrared, as $\t\to 0$,
\beq
dY_5^2\qquad \underbrace{\longrightarrow}_{\t\to 0}\qquad \frac 16 \ep^{4/3} 2^{1/3} 3^{2/3} H_{KS}^{1/2}\bigg|_{\t\to 0}\
\biggl[\ \left(\frac 12 g_5^2+g_3^2+g_4^2\right)+\frac {\t^2}{4}\ \left(g_1^2+g_2^2\right)\ \biggr]\,.
\label{ksir}
\eeq
In \eqref{ksir} we can easily identify the $2-$ and $3-$cycles:
\beq\begin{aligned}
&2-{\rm cycle}:\ g_1^2+g_2^2\bigg|_{\psi=0,\theta_2=-\theta_1,\phi_2=-\phi_1}=2\
\biggl(\ (d\theta_1)^2+ \sin^2\theta_1\ (d\phi_1)^2\ \biggr)\,,\\
&3-{\rm cycle}:\ \frac
12 g_5^2+g_3^2+g_4^2\bigg|_{\phi_2=\theta_2=0,\psi=\xi_1+\xi_2,\phi_1=\xi_2-\xi_1,\theta_1=2\eta}
=\\
&\qquad\qquad\qquad\qquad\qquad\qquad  2\biggl(\ (d\eta)^2+\sin^2\eta\ (d\xi_1)^2+\cos^2\eta\ (d\xi_2)^2
\
\biggr)\,.
\label{kscycles}\end{aligned}
\eeq
In the ultraviolet, as $\t\to \infty$,
\beq
\frac{dY_5^2}{\Om_2^2}\qquad \underbrace{\longrightarrow}_{\t\to \infty}\qquad
6\biggl[\ \frac 19 g_5^2+\frac16 \left(g_3^2+g_4^2\right)+\frac16 \left(g_1^2+g_2^2\right)
  \ \biggr]\,,
\label{ksuv1}
\eeq
which is the standard metric on the coset $T^{1,1}=\frac{SU(2)\times SU(2)}{U(1)}$ of radius $R_{T^{1,1}}=\sqrt{6}$.
As in \eqref{kscycles} we can identify the $2-$ and $3-$cycles
\beq
\frac{dY_5^2}{\Om_2^2}\qquad \underbrace{\longrightarrow}_{\t\to \infty}\qquad
\underbrace{\left[ \delta_\infty\ \frac 12 g_5^2 +g_3^2+g_4^2\right]}_{3-{\rm cycle}}
+\underbrace{\left[g_1^2+g_2^2\right]}_{2-{\rm cycle}}\,,\qquad \delta_\infty=\frac 43\,.
\label{ksuv2}
\eeq
%
%HERE
%
Note that at the tip of the deformed conifold, $Y_5$, the $3$-cycle is a
round $S^3$, and the $2$-cycle is a collapsed $S^2$. Because $\delta_\infty\ne 1$,
the $3-$cycle is a deformed $S^3$ as $\t\to \infty$. 
It is convenient to (formally) assign radii for the $2-$ and $3$-cycles as:
\beq
R_{3}\equiv \sqrt{2}\Om_2\,,\qquad R_{2}\equiv \sqrt{2}\Om_3\,.
\label{rad}
\eeq
We can now introduce  parameters
\beq
\xi\equiv 1-\frac{R_2^2}{R_3^2}=1-\frac{\Om_3^2}{\Om_2^2}\,,\qquad \delta\equiv \frac{2\Om_1^2}{\Om_2^2}\,,
\label{xdefxi}
\eeq
where $\xi$ is the order parameter for the chiral symmetry breaking $U(1)\to \zet_2$ in the dual gauge theory
and geometrically encodes the relative sizes of the compact cycles on the deformed conifold;
$\delta$ encodes the deformation of the $3-$cycle relative to a round $S^3$.

To facilitate the comparison of the extremal KS geometry \eqref{ks1} with that of the
KS BH we introduce a new radial coordinate $\r$ as
\begin{equation}
\frac{(d\r)^2}{\r^4}\equiv g_{{\bm x}{\bm x}}\ g_{\t\t}\ (d\t)^2=(w_{1,KS}(\t))^2 (d\t)^2\,.
\label{rrho}
\end{equation}
Note that the asymptotic boundary $\t\to\infty$ corresponds to $\r\to 0$.
With
\beq
e^{-\t}\equiv \frac{3\sqrt{6}\ep^2}{8}\ z^3\,,
\eeq
we find from \eqref{rrho}
\beq
z=\r \left(1+\r^6\ep^4 \left(\frac{27}{800}
+\frac{27}{80}\ \ln 3-\frac{9}{16}\ \ln2+\frac{27}{40}\ \ln\left(\rho\ep^{2/3}\right)\right)+
{\cal O}(\r^{12}\ep^8\ln^2\left(\rho\ep^{2/3}\right))\right)\,,
\eeq
leading to
\beq
K_{1,KS}\equiv 12 P\ h_{1,KS}=P^2 g_s
\biggl( -\ln 3 +\frac 53\ln 2-\frac 43\ln\ep-\frac 23-2\ln \rho+{\cal O}(\r^3\ep^2\ln\left(\r\ep^{2/3}
\right)
\biggr) \,.
\label{ask1}
\eeq
In Appendix~\ref{appendixB} we will use \eqref{ask1} to relate $\Lambda$ and $\ep$.

We conclude this section relating the Einstein frame KS solution \eqref{ks2} with the
String frame solution reviewed in \cite{Herzog:2001xk}.
 Substituting explicit expressions in \eqref{ks2} for $H_{KS}$ we find
\beq
  H_{KS}\equiv \frac{2^{2/3}g_s}{\epsilon^{8/3}}\ \left(\frac{2P}{9}\right)^2\ I(\t)=\frac{2^{2/3}g_s}{\epsilon^{8/3}}\ \left(\frac{2P}{9}\right)^2\ \int_{\t}^\infty dx\ \frac{x \coth x-1}{\sinh^2 x}
  (\sinh(2x)-2x)^{1/3}\,.
  \label{hksexp}
  \eeq
Note that the integral $I(\t)$ in \eqref{hksexp} is exactly the same as in eq.(66) of \cite{Herzog:2001xk}. In the limit $\t\to 0$,
\beq
  I(\t\to 0)= I(0)\qquad \Longrightarrow\qquad  H_{KS}(\t\to 0)=\frac{2^{2/3}g_s}{\epsilon^{8/3}} \left(\frac{2P}{9}\right)^2\ I(0)\,.
\label{i0def}
  \eeq
  We already established  (from the quantization of 3-form flux \eqref{defm}) a relation between
  $P$ and $M$ \eqref{ks3}, so
  \beq
  H_{KS}(\t\to 0)=\frac{2^{2/3}M^2(\a')^2g_s I(0)}{\epsilon^{8/3}}\,.
  \eeq
The Einstein frame metric as $\t\to 0$ is then
\begin{equation}
  \begin{aligned}
  ds_{10}^2&\bigg|_{Einstein}^{KS}=\frac{\epsilon^{4/3}}{2^{1/3}(I(0))^{1/2}g_s^{1/2} M\a'}dx_ndx_n
  \\
  &\qquad\qquad\qquad +(I(0))^{1/2}6^{-1/3}g_s^{1/2} M\a'\biggl\{
  \frac 12(d\t)^2+\frac 12 g_5^2+g_3^2+g_4^2+\frac 14\t^2[g_1^2+g_2^2]\biggr\}\,.
\end{aligned}
  \end{equation}
Since
  \[
  ds_{10}^2\bigg|_{string}=g_s^{1/2}\  ds_{10}^2\bigg|_{Einstein}\,,
  \]
we have
   \begin{equation}
  \begin{aligned}
  ds_{10}^2&\bigg|_{string}^{KS}=\frac{\epsilon^{4/3}}{2^{1/3}(I(0))^{1/2} M\a'}dx_ndx_n
  \\
  &\qquad\qquad\qquad+(I(0))^{1/2}6^{-1/3}g_s M\a'\biggl\{
  \frac 12(d\t)^2+\frac 12 g_5^2+g_3^2+g_4^2+\frac 14\t^2[g_1^2+g_2^2]\biggr\}\,.
\label{us}
  \end{aligned}
   \end{equation}
   Note that
   \beq
   H_{KS,string}=\frac{1}{g_s}\ H_{KS}\,.
   \eeq
 Our expression \eqref{us} differs from eq.(68) of \cite{Herzog:2001xk} (HKO):
     \begin{equation}
  \begin{aligned}
  ds_{10}^2&\bigg|_{HKO}^{KS}=\frac{\epsilon^{4/3}}{2^{1/3}I(0)^{1/2} g_sM\a'}dx_ndx_n
  \\&\qquad\qquad\qquad +I(0)^{1/2}6^{-1/3}g_s M\a'\biggl\{
  \frac 12(d\t)^2+\frac 12 g_5^2+g_3^2+g_4^2+\frac 14\t^2[g_1^2+g_2^2]\biggr\}\,;
\label{hko}
  \end{aligned}
     \end{equation}
    while the  warp factors  in $ds_6^2$ agree,
    there is a typo in the $dx_ndx_n$ warp factor.

\section{The parameters $\Lambda$, $\xi$ and $\delta$ of the KS black hole}\label{appendixB}

We recall first
relevant details of the KS BH constructed numerically in \cite{Buchel:2018bzp}.
The background geometry is 
\beq
ds_{10}^2 =H^{-1/2}\biggl(-(1-x)^2 dt^2+(d{\bm x})^2\biggr)+g_{xx}\ dx^2+\Om_1^2 g_5^2
+\Om_2^2 \left[g_3^2+g_4^2\right]+\Om_3^2 \left[g_1^2+g_2^2\right]\,,
\label{bh1}
\eeq
where the radial coordinate is $x\in [0,1]$, with
\beq\begin{aligned}
h_1=&\frac {1}{12P}\ {K_1}\,,\qquad h_2=\frac{P}{18}\ K_2\,,\qquad 
h_3=\frac {1}{12P}\ K_3\,,\qquad H=(2x-x^2) h\,,\\
\Om_1=&\frac 13 f_c^{1/2} h^{1/4}\,,\qquad \Om_2=\frac {1}{\sqrt{6}} f_a^{1/2} h^{1/4}\,,\qquad 
\Om_3=\frac {1}{\sqrt{6}} f_b^{1/2} h^{1/4}\,.
\end{aligned}
\label{bh2}
\eeq
Asymptotically near the boundary ($x\to 0$),
\beq\begin{aligned}
K_1=&P^2 g_s\biggl[ k_s-\frac12 P^2 g_0\ \ln x+\sum_{n=3}^\infty\sum_k\ k_{1nk}\ x^{n/4}\ \ln^k x\biggr]\,,
\end{aligned}
\label{bh31}
\eeq
\beq\begin{aligned}
K_2=&1+\sum_{n=3}^\infty\sum_k\ k_{2nk}\ x^{n/4}\ \ln^k x\,,
\end{aligned}
\label{bh32}
\eeq
\beq\begin{aligned}
K_3=&P^2 g_s\biggl[ k_s-\frac12 P^2 g_0\ \ln x+\sum_{n=3}^\infty\sum_k\ k_{3nk}\ x^{n/4}\ \ln^k x\biggr]\,,
\end{aligned}
\label{bh33}
\eeq
\beq\begin{aligned}
f_a=&a_0\biggl[1+\sum_{n=3}^\infty\sum_k\ f_{ank}\ x^{n/4}\ \ln^k x\biggr]\,,
\end{aligned}
\label{bh34}
\eeq
\beq\begin{aligned}
f_b=&a_0\biggl[1+\sum_{n=3}^\infty\sum_k\ f_{bnk}\ x^{n/4}\ \ln^k x\biggr]\,,
\end{aligned}
\label{bh35}
\eeq
\beq\begin{aligned}
f_c=&a_0\biggl[1+\sum_{n=2}^\infty\sum_k\ f_{cnk}\ x^{n/4}\ \ln^k x\biggr]\,,
\end{aligned}
\label{bh36}
\eeq
\beq\begin{aligned}
h=&\frac{P^2g_s}{a_0^2}\biggl[
  \left(\frac 18+\frac {k_s}{4}\right)-\frac  {1}{8}\ \ln x+\sum_{n=2}^\infty\sum_k\ h_{nk}\ x^{n/4}\ \ln^k x\biggr]\,,\\
g=&g_s\biggl[\ 1+\sum_{n=2}^\infty\sum_k\ g_{nk}\ x^{n/2}\ \ln^k x \ \biggr]\,,
\end{aligned}
\label{bh37}
\eeq
and the solution is characterized by four microscopic parameters $\{P^2\,, g_s\,, a_0\,, k_s\}$ and seven
expectation values $\{{ df_0}\equiv(f_{a30}-f_{b30})/2,
dk_{10}\equiv (k_{130}-k_{330})/2,f_{a40},g_{40},f_{a60},k_{270},f_{a80}\}$.
Asymptotically near the regular horizon (as $y\equiv 1-x\to 0$) the KS BH solution
\beq\begin{aligned}
&K_i=P^2 g_s\sum_{n=0}^\infty k_{ihn}\ y^{2n}\,,\qquad i=1,3\,,\qquad K_2=\sum_{n=0}^\infty k_{2hn}\ y^{2n}\,,\\
&f_{\alpha}=a_0\ \sum_{n=0}^\infty f_{\a hn}\ y^{2n}\,,\qquad \a=a,b,c\,,\\
&h=\frac{P^2g_s}{a_0^2}\ \sum_{n=0}^\infty h_{hn}\ y^{2n}\,,\qquad g=g_s\ \sum_{n=0}^\infty g_{hn}\ y^{2n}\,,
\end{aligned}
\label{bh4}
\eeq
is characterized by the ``nice'' parameters $\{k_{1h0}\,, k_{2h0}\,, k_{3h0}\,,f_{ah0}\,,  f_{bh0}\,,  f_{ch0}\,, f_{ch1}\,,
h_{h0}\,, g_{h0}\}$.

The microscopic parameters
$k_s$ and $a_0$ determine the strong-coupling scale $\Lambda$ of the theory via
\beq
k_s=\frac 12 \ln\left(\frac{a_0^2}{\Lambda^4}\right)\,,
\label{bh5}
\eeq
and the entropy and the energy densities are given by 
\beq\begin{aligned}
16\pi G_5\ \frac{s}{\Lambda^3}=&4\pi\ P g_s^{1/2}\ e^{3 k_s/2}\  h_{h0}^{1/2}\ f_{ch0}^{1/2}\ f_{ah0}\ f_{bh0}\,,\\
16\pi G_5\ \frac{\cal E}{\Lambda^4}=&\ e^{2 k_s}\ (3 -12\ f_{a40})\,,
\end{aligned}
\label{bh6}
\eeq
where
\beq
\frac{1}{16\pi G_5}=\frac{{\rm vol}(T^{1,1})}{16\pi G_{10}}=\frac{16\pi^3}{27}\ \frac{1}{16\pi G_{10}}=\frac{1}{216\pi^4 (\a')^4} \,.
\label{bh7}
\eeq

\begin{figure}[th]
\centering
\includegraphics[width=.48\textwidth]{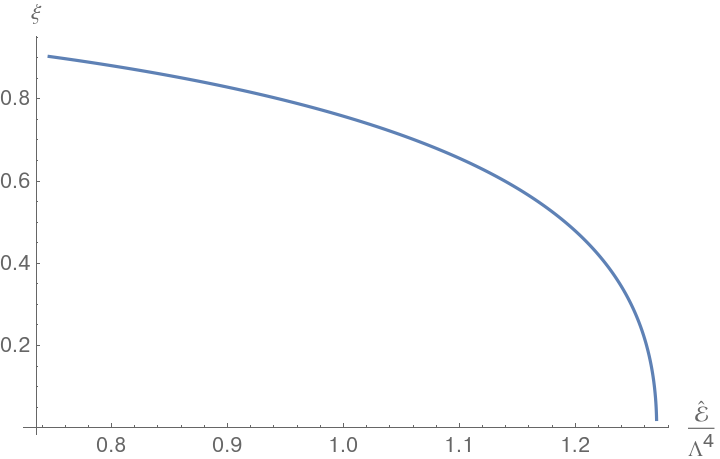}
\hspace{0.5cm}
\includegraphics[width=.46\textwidth]{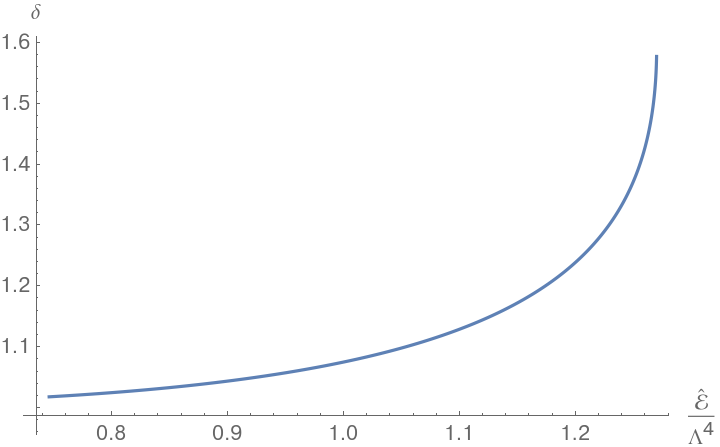}
  \caption{
{\bf Left Panel:}  Order parameter $\xi$ for the chiral symmetry breaking in KS BH.
{\bf Right Panel:}  Deformation parameter $\delta$ of the
3-cycle relative to a round $S^3$  in KS BH.} \label{figure1}
\end{figure}   
We now proceed to relate $\Lambda$ to the conifold deformation parameter
$\ep$, as we did for the supersymmetric solution in Appendix \ref{appendixA}. We first identify the
``holographic'' radial coordinate, $\rho$, exactly as in equation \eqref{rrho}:\footnote{To avoid confusion, we should remember that $g_{{\bm x}{\bm x}}$ is the metric along the 3 spatial directions, denoted generically by ${\bm x}$, and $g_{xx}$ is the metric along the radial direction, denoted by $x$.}
\beq
\frac{(d\r)^2}{\r^4}\equiv g_{{\bm x}{\bm x}}\ g_{xx}\ (dx)^2\,.
\label{bh8}
\eeq
Using the expansions in \eqref{bh31}-\eqref{bh37} we find
\beq
x^{1/4}=  2^{-1/4} \rho a_0^{1/2} \biggl(
1+\rho^4 a_0^2\ \left(\frac 14 f_{a40}-\frac{5}{48}\right)
+\rho^6 a_0^3\ \frac{3\sqrt{2}}{400}   df_0^2+{\cal O}\left(\rho^8 a_0^4\ln^2
\left(\rho a_0^{1/2}\right)\right)\biggr)\,,
\label{bh9}
\eeq
and using equation \eqref{bh5} we can express $K_1$ in terms of the ``holographic'' radial variable 
\beq
K_1=P^2g_s\biggl( -2\ln\Lambda+\frac 12\ln2-2\ln\rho + {\cal O}
\left(\rho^3 a_0^{3/2}\ln
\left(\rho a_0^{1/2}\right)\right)\biggr)\,.
\label{bh10}
\eeq
Comparing $K_1$ from \eqref{bh10} and
$K_{1,KS}$ from \eqref{ask1} we identify
\beq
\Lambda=\frac{3^{1/2}e^{1/3}}{2^{7/12}}\ \ep^{2/3}\,.
\label{lres}
\eeq
Given the numerical solution of the KS BH it is straightforward to
evaluate the parameters $\xi$  and $\delta$ \eqref{xdefxi} at the horizon
\beq
\xi_{horizon}=1-\frac{f_{bh0}}{f_{ah0}}\,,\qquad
\delta_{horizon}=\frac{4f_{ch0}}{3f_{ah0}}\,.
\label{xideltares}
\eeq
Numerical results for these parameters are presented in Fig.~\ref{figure1}.

%%%%%%%%%%%%%%%%%%%%%%%%%%%%%%%%%%%%%%%%%%%%%%%%%%%%%%%%%%%
%\newpage

\providecommand{\href}[2]{#2}\begingroup\raggedright\endgroup


\begin{thebibliography}{10}


%\cite{Bena:2018fqc}
\bibitem{Bena:2018fqc}
  I.~Bena, E.~Dudas, M.~Gra\~na and S.~L\"ust,
  ``Uplifting Runaways,''
  Fortsch.\ Phys.\  {\bf 67} (2019) no.1-2,  1800100
  %doi:10.1002/prop.201800100
  [arXiv:1809.06861 [hep-th]].
  %%CITATION = doi:10.1002/prop.201800100;%%
  %29 citations counted in INSPIRE as of 27 Aug 2019

%\cite{Buchel:2018bzp}
\bibitem{Buchel:2018bzp}
  A.~Buchel,
  ``Klebanov-Strassler black hole,''
  JHEP {\bf 1901} (2019) 207
  %doi:10.1007/JHEP01(2019)207
  [arXiv:1809.08484 [hep-th]].
  %%CITATION = doi:10.1007/JHEP01(2019)207;%%
  %1 citations counted in INSPIRE as of 27 Aug 2019

%\cite{Kachru:2003aw}
\bibitem{Kachru:2003aw} 
  S.~Kachru, R.~Kallosh, A.~D.~Linde and S.~P.~Trivedi,
  ``De Sitter vacua in string theory,''
  Phys.\ Rev.\ D {\bf 68}, 046005 (2003)
  %doi:10.1103/PhysRevD.68.046005
  [hep-th/0301240].
  %%CITATION = doi:10.1103/PhysRevD.68.046005;%%
  %2726 citations counted in INSPIRE as of 13 Sep 2019

%\cite{Klebanov:2000hb}
\bibitem{Klebanov:2000hb}
  I.~R.~Klebanov and M.~J.~Strassler,
  ``Supergravity and a confining gauge theory: Duality cascades and chi SB resolution of naked singularities,''
  JHEP {\bf 0008} (2000) 052
  %doi:10.1088/1126-6708/2000/08/052
  [hep-th/0007191].
  %%CITATION = doi:10.1088/1126-6708/2000/08/052;%%
  %1638 citations counted in INSPIRE as of 27 Aug 2019

%\cite{Klebanov:2000nc}
\bibitem{Klebanov:2000nc} 
  I.~R.~Klebanov and A.~A.~Tseytlin,
  ``Gravity duals of supersymmetric SU(N) x SU(N+M) gauge theories,''
  Nucl.\ Phys.\ B {\bf 578}, 123 (2000)
  %doi:10.1016/S0550-3213(00)00206-6
  [hep-th/0002159].
  %%CITATION = doi:10.1016/S0550-3213(00)00206-6;%%
  %518 citations counted in INSPIRE as of 13 Sep 2019
  


%\cite{Buchel:2010wp}
\bibitem{Buchel:2010wp}
  A.~Buchel,
  ``Chiral symmetry breaking in cascading gauge theory plasma,''
  Nucl.\ Phys.\ B {\bf 847} (2011) 297
  %doi:10.1016/j.nuclphysb.2011.01.031
  [arXiv:1012.2404 [hep-th]].
  %%CITATION = doi:10.1016/j.nuclphysb.2011.01.031;%%
  %15 citations counted in INSPIRE as of 27 Aug 2019

%\cite{Aharony:2007vg}
\bibitem{Aharony:2007vg} 
  O.~Aharony, A.~Buchel and P.~Kerner,
  ``The Black hole in the throat: Thermodynamics of strongly coupled cascading gauge theories,''
  Phys.\ Rev.\ D {\bf 76}, 086005 (2007)
  %doi:10.1103/PhysRevD.76.086005
  [arXiv:0706.1768 [hep-th]].
  %%CITATION = doi:10.1103/PhysRevD.76.086005;%%
  %70 citations counted in INSPIRE as of 13 Sep 2019

%\cite{Bena:2014jaa}
\bibitem{Bena:2014jaa} 
  I.~Bena, M.~Grana, S.~Kuperstein and S.~Massai,
  ``Giant Tachyons in the Landscape,''
  JHEP {\bf 1502}, 146 (2015)
  %doi:10.1007/JHEP02(2015)146
  [arXiv:1410.7776 [hep-th]].
  %%CITATION = doi:10.1007/JHEP02(2015)146;%%
  %64 citations counted in INSPIRE as of 13 Sep 2019
%\cite{Bena:2016fqp}

\bibitem{Bena:2016fqp} 
  I.~Bena, J.~Blaback and D.~Turton,
  ``Loop corrections to the antibrane potential,''
  JHEP {\bf 1607}, 132 (2016)
  %doi:10.1007/JHEP07(2016)132
  [arXiv:1602.05959 [hep-th]].
  %%CITATION = doi:10.1007/JHEP07(2016)132;%%
  %23 citations counted in INSPIRE as of 13 Sep 2019

%\cite{Douglas:2008jx}
\bibitem{Douglas:2008jx} 
  M.~R.~Douglas and G.~Torroba,
  ``Kinetic terms in warped compactifications,''
  JHEP {\bf 0905}, 013 (2009)
  %doi:10.1088/1126-6708/2009/05/013
  [arXiv:0805.3700 [hep-th]].
  %%CITATION = doi:10.1088/1126-6708/2009/05/013;%%
  %67 citations counted in INSPIRE as of 13 Sep 2019
  
%\cite{Bena:2018fqc}
\bibitem{tadpole-paper}
  I.~Bena, J.~Blaback, M.~Gra\~na and S.~L\"ust,
  ``The Tadpole Problem,''
  [arXiv:1910.XXXXX [hep-th]].

%\cite{Blumenhagen:2019qcg}
\bibitem{Blumenhagen:2019qcg}
  R.~Blumenhagen, D.~Kl\"awer and L.~Schlechter,
  ``Swampland Variations on a Theme by KKLT,''
  JHEP {\bf 1905} (2019) 152
  %doi:10.1007/JHEP05(2019)152
  [arXiv:1902.07724 [hep-th]].
  %%CITATION = doi:10.1007/JHEP05(2019)152;%%
  %12 citations counted in INSPIRE as of 27 Aug 2019

%\cite{Herzog:2001xk}
\bibitem{Herzog:2001xk} 
  C.~P.~Herzog, I.~R.~Klebanov and P.~Ouyang,
  ``Remarks on the warped deformed conifold,''
  hep-th/0108101.
  %%CITATION = HEP-TH/0108101;%%
  %154 citations counted in INSPIRE as of 19 Sep 2019


%\cite{Buchel:2013dla}
\bibitem{Buchel:2013dla} 
  A.~Buchel and D.~A.~Galante,
  ``Cascading gauge theory on $dS_4$ and String Theory landscape,''
  Nucl.\ Phys.\ B {\bf 883}, 107 (2014)
  %doi:10.1016/j.nuclphysb.2014.03.022
  [arXiv:1310.1372 [hep-th]].
  %%CITATION = doi:10.1016/j.nuclphysb.2014.03.022;%%
  %18 citations counted in INSPIRE as of 19 Sep 2019


\end{thebibliography}
\end{document}